\documentclass[twocolumn,english,aps,prc,groupedaddress,amsmath,amssymb,showpacs]{revtex4}
\usepackage[T1]{fontenc}
\usepackage[latin9]{inputenc}
\usepackage{amsmath}
\usepackage{amssymb}

\makeatletter

\providecommand{\tabularnewline}{\\}

\@ifundefined{textcolor}{}
{%
 \definecolor{BLACK}{gray}{0}
 \definecolor{WHITE}{gray}{1}
 \definecolor{RED}{rgb}{1,0,0}
 \definecolor{GREEN}{rgb}{0,1,0}
 \definecolor{BLUE}{rgb}{0,0,1}
 \definecolor{CYAN}{cmyk}{1,0,0,0}
 \definecolor{MAGENTA}{cmyk}{0,1,0,0}
 \definecolor{YELLOW}{cmyk}{0,0,1,0}
 }

\usepackage{babel}

\makeatother

\usepackage{babel}
\begin{document}

\title{Odd-$J$ Pairing in Nuclei}

\author{L. Zamick}

\author{A. Escuderos}

\affiliation{Department of Physics and Astronomy, Rutgers University, Piscataway,
NJ 08854, USA}
\begin{abstract}
We point out a simplicity that arises when we use an interaction in
which only an energy with odd $J$ is non-zero. The emphasis is on
$J=J_{\text{max}}$ and in particular $J=9^{+}$ in the $g_{9/2}$
shell. It is noted that high overlaps can be deceptive. In many cases
a single set of unitary 9-$j$ coefficients gives either an exact
or a surprisingly good approximation to the wave function of a non-degenerate
state. The many degeneracies that occur in these calculations are
discussed and explained. As a counterpoint, we compare the results
with an interaction in which both the $J=0$ and $J=J_{\text{max}}$
two-body matrix elements are equal (and attractive). Comparisons with
a more realistic interaction are also made. 
\end{abstract}
\maketitle

\section{Introduction}

The purpose of this work is to study the properties of a very simple
interaction in a single-$j$-shell model space of neutrons and protons.
A single proton--neutron ($pn$) pair in this space can have a total
angular momentum from $J=0$ to $J=J_{\text{max}}=2j$. The even-$J$
states have isospin $T=1$, i.e. they are members of an isotriplet---there
are analog states of two neutrons and of two protons with these even
angular momenta. The odd-$J$ states have isospin $T=0$, i.e. isosinglet---in
this model space, they only are present in the $pn$ system.

There has been much study in journals and in textbooks of the $J=0$,
$T=1$ {}``pairing interacton'', i.e. where only $J=0$ two-body
matrix elements are non-zero (and attractive)~\cite{r52,f52,rt53,st63,t93}.
Although the pairing interaction is not realistic, the consequences
of these studies have yielded results whose importance was well beyond
expectation. Examples are seniority classifications, reduced isospin
and indeed these studies can be regarded as precursors to the BCS
theory in condensed matter theory. In turn, the BCS theory is important
in the nuclear context for explaining moments of inertia of deformed
states.

In the $pn$ system, it is known that not only the $J=0$ two-body
matrix element lies low but also $J=1$ and $J=J_{\text{max}}=2j$.
Therefore, as a counterpoint to the $J=0$ pairing interaction, we
will here consider $J=J_{\text{max}}$ pairing interaction, for which
all two-body matrix elements are equal to zero except for the $J=J_{\text{max}}$
two-body matrix element, which is attractive. Such an interaction
only acts between a neutron and a proton, not between two identical
particles. This study will reveal some surprising results.

It should be noted that there has been a recent flurry of interest
in $J_{\text{odd}}$ pairing and in particular the case where $J_{\text{odd}}$
is equal to $J_{\text{max}}$. In a recent work, C. Qi et al.~\cite{qetal11}
proposed a wave function for $^{96}$Cd in which each of two $pn$
pairs couple to maximum angular momentum, which for the $g_{9/2}$
shell is $J=9$. They start their approach with basis states of two
protons coupling to $J_{p}$ and two neutrons to $J_{n}$. In this
basis, the effect of the Pauli principle is clear. Once one has antisymmetric
wave functions of two protons, which is achieved by limiting the angular
momenta to even $J_{p}$, and likewise the two neutrons to even $J_{n}$,
one has a wave function which satisfies the Pauli principle. Then
they consider the wave function in terms of $pn$ pairs, in which
one $pn$ pair couples to $J_{1}$ and the other to $J_{2}$. For
this wave function, $|[pn(J_{1})\, pn(J_{2})]^{I}\rangle$, the antisymmetrization
looks at first complicated. But the authors assure us that {}``The
overlap matrix automatically takes into account the Pauli principle''.
And this overlap matrix between the two previous bases can be expressed
in terms of a 9-j symbol as follows (see Eq.~(2) of Ref.~\cite{qetal11}):
\begin{eqnarray}
\lefteqn{\langle[pn(J_{1})pn(J_{2}){]}^{I}|[pp(J_{p})nn(J_{n})]^{I}\rangle=}\nonumber \\
 & = & \frac{-2}{\sqrt{N_{J_{1}J_{2}}}}\hat{J_{1}}\hat{J_{2}}\hat{J_{p}}\hat{J_{n}}\begin{Bmatrix}j & j & J_{p}\\
j & j & J_{n}\\
J_{1} & J_{2} & I
\end{Bmatrix}\,,
\end{eqnarray}
 We will here test the ansatz that these overlaps make up the components
of an exact or approximate eigenfunction of an odd-$J$ pairing interaction.
In this work we use the symbol $I$ for the total angular momentum
of the state and $J$ otherwise.

We here note that work we previously did shows that certain $U9$-$j$
coefficients form components of the $J=0^{+}$ ground state wave function
of a $J_{\text{odd}}$ pairing interaction. We here expand on this
work. We refer to a previous work by E.~Moya de Guerra et al.~\cite{mrzs03}
and explicitly to Eqs. (70) and (74). In that work we describe the
wave functions in a $|[pp(J_{p})\, nn(J_{n})]^{I}\rangle$ basis.
The Pauli principle is easily satisfied by constraining $J_{p}$ and
$J_{n}$ to be even. Our previous example was $^{44}$Ti, but the
same mathematics holds for $^{96}$Cd. We previously considered various
schematic interactions as well and the more realistic MBZ interaction
taken from experiment~\cite{bmz63}, for which detailed wave funtions
were subsequently published in the archives~\cite{ezb05} (with some
modification of the two-body matrix elements).

\section{Results\label{sec:results}}

We now proceed to the calculations. We first define a unitary 9-$j$
symbol $U9$-$j$ as the follows: 
\begin{eqnarray}
\lefteqn{|[j_{1}j_{2}(J_{12})j_{3}j_{4}(J_{34})]^{I}\rangle=}\nonumber \\
 & = & \sum_{J_{13}J_{24}}\langle(j_{1}j_{2})^{J_{12}}(j_{3}j_{4})^{J_{34}}|(j_{1}j_{3})^{J_{13}}(j_{2}j_{4})^{J_{24}}\rangle^{I}\times\nonumber \\
 &  & \times|[j_{1}j_{3}(J_{13})j_{2}j_{4}(J_{24})]^{I}\rangle
\end{eqnarray}

We will be extensively using the following two relations for the $U9$-$j$'s
(taken from \textit{Nuclear Shell Theory}, p.~516, by de-Shalit and
Talmi~\cite{st63}): 
\begin{eqnarray}
\lefteqn{\sum_{J_{13}J_{24}}{\langle(jj)^{J_{a}}(jj)^{J_{b}}|(jj)^{J_{13}}(jj)^{J_{24}}\rangle^{I}}\times}\nonumber \\
 &  & \times\langle(jj)^{J_{c}}(jj)^{J_{d}}|(jj)^{J_{13}}(jj)^{J_{24}}\rangle^{I}=\nonumber \\
 & = & \delta_{a,c}\delta_{b,d}\label{eq:orth}\\
\lefteqn{\sum_{J_{13}J_{24}}{(-1)^{s}\langle(jj)^{J_{12}}(jj)^{J_{34}}|(jj)^{J_{13}}(jj)^{J_{24}}\rangle^{I}}\times}\nonumber \\
 &  & \times\langle(jj)^{J_{13}}(jj)^{J_{24}}|(jj)^{J_{14}}(jj)^{J_{23}}\rangle^{I}=\nonumber \\
 & = & \langle(jj)^{J_{12}}(jj)^{J_{34}}|(jj)^{J_{14}}(jj)^{J_{23}}\rangle^{I}\,,\label{eq:u9j-rel2}
\end{eqnarray}
 where $s=J_{24}+J_{23}-J_{34}-1$.

Let us first consider the $I=0^{+}$ states in $^{96}$Cd (or $^{44}$Ti).
For most interactions, the diagonalization is a fairly complicated
procedure. However for certain interactions it is much easier. For
example, the interaction used in Ref.~\cite{mrzs03} was one in which
all two-body matrix elements were set equal to zero except for the
$J=1$, $T=0$ two-body matrix element. In such a case, the matrix
element of the secular four-particle Hamiltonian factorizes (the same
result holds for any odd-$J$ interaction). This is the key point.
The basis states are $|[pp(J_{p})\, nn(J_{n})]^{I=0}\rangle$, where
$J_{n}=J_{p}$, and the Hamiltonian takes the following form: 
\begin{equation}
H_{J_{p},J_{p'}}=V(J_{\text{odd}})f(J_{p})f(J_{p'})\,,
\end{equation}
 where $f(J_{p})$ is twice the $U9$-$j$ symbol: 
\begin{eqnarray}
f(J_{p}) & = & 2\langle(jj)^{J_{p}}(jj)^{J_{p}}|(jj)^{J_{\text{odd}}}(jj)^{J_{\text{odd}}}\rangle^{I=0}=\nonumber \\
 & = & 2(2J_{p}+1)(2J_{\text{odd}}+1)\begin{Bmatrix}j & j & J_{p}\\
j & j & J_{p}\\
J_{\text{odd}} & J_{\text{odd}} & 0
\end{Bmatrix}
\end{eqnarray}

If we write the wave function as $\sum X_{J_{p}J_{p}}|[pp(J_{p})nn(J_{p})]^{I=0}\rangle$
(as in Ref.~\cite{qetal11}), then it was shown in Ref.~\cite{mrzs03}
that $X_{J_{p}J_{p}}$ is proportional to $f(J_{p})$. The other $I=0^{+}$
eigenstates are degenerate and, if $V(J_{\text{odd}})$ is negative,
they are at higher energies. In other words, what we have shown in
Ref.~\cite{mrzs03} is that the wave-function components $X_{J_{p}J_{p}}$
of the lowest $I=0^{+}$ state are proportional to the overlap factor
of Ref.~\cite{qetal11}; alternately, they are equal within a normalization
to the $U9$-$j$ coefficients.

The eigenvalue is given by 
\begin{equation}
E(I=0^{+})=V(J_{\text{odd}})|\sum f(J_{p})X_{J_{p}J_{p}}|^{2}
\end{equation}
 Note that our very simple interactions are charge independent. This
means that the lowest (non-degenerate) $I=0^{+}$ state has good isospin,
presumably $T=0$. It is amusing that we can assign the isospin quantum
number to a wave function with $U9$-$j$ coefficients.

In Table~\ref{tab:wf} we present the wave functions for the following
interactions: 
\begin{description}
\item [{{CCGI:}}] A realistic interaction fit as well as possible to
experiment (see Ref.~\cite{ccgi12}). 
\item [{{E(0):}}] $V(0)=-2.0000$~MeV; all other matrix elements are
zero. 
\item [{{E(9):}}] $V(9)=-2.0000$~MeV; all other matrix elements are
zero. 
\item [{{E(0,9):}}] $V(9)=V(0)=-2.0000$~MeV; all other matrix elements
are zero. 
\item [{{E(1):}}] $V(1)=-2.0000$~MeV; all other matrix elements are
zero. 
\end{description}
\begin{table}[htb]
 \caption{\label{tab:wf} Wave Functions of the $I=0^{+}$ ground state of $^{96}$Cd
for various interactions.}

\begin{ruledtabular} %
\begin{tabular}{cccccc}
 & CCGI  & E(0)  & E(9)  & E(0,9)  & E(1)\tabularnewline
\hline 
$X_{00}$  & 0.7725  & 0.8563  & 0.6164  & 0.8103  & 0.2903 \tabularnewline
$X_{22}$  & 0.5280  & 0.1741  & 0.7518  & 0.4814  & 0.5704 \tabularnewline
$X_{44}$  & 0.2915  & 0.2335  & 0.2385  & 0.2514  & 0.5190 \tabularnewline
$X_{66}$  & 0.1704  & 0.2807  & 0.0233  & 0.1718  & 0.1586 \tabularnewline
$X_{88}$  & 0.1020  & 0.3210  & 0.0005  & 0.1831  & $-0.5540$ \tabularnewline
\end{tabular}\end{ruledtabular} 
\end{table}

We consider the second column (CCGI) as the realistic interaction
to which the other interactions should be compared. The simplest thing
we can do is give the overlaps of the above interactions with CCGI.
They are respectively 0.9020, 0.9451, 0.9944, and 0.6484. We find
that E(9) gives higher overlap than the much studied E(0) pairing
interaction and a much higher overlap than E(1). This might lead one
to believe that the idea of $J=9^{+}$ pairing is a valid concept.
But overlaps can be deceiving. We also present E(0,9), where the only
non-vanishing matrix elements are for $J=9^{+}$ and $0^{+}$, both
set to $-2.0000$~MeV. Now the overlap is even higher---0.9944. This
might not be startlingly different than 0.9467, but let us now look
at the energies of the lowest even-$I$ states in Table~\ref{tab:yrast}.
They are given respectively for interactions CCGI, E(9), and E(0,9).
The results of the second column (CCGI) were previously given~\cite{ze12}
and the point was made that the $I=16^{+}$ state is isomeric since
it lies below the lowest $14^{+}$ and $15^{+}$ states. This is in
agreement with experiment~\cite{setal11}.

\begin{table}[htb]
 \caption{\label{tab:yrast} Calculated spectra of yrast even-$I$ states in
$^{96}$Cd for above mentioned interactions.}

\begin{ruledtabular} %
\begin{tabular}{cccc}
$I^{\pi}$  & CCGI  & E(9)  & E(0,9)\tabularnewline
\hline 
$0^{+}$  & 0.0000  & 1.0587  & 0.0000 \tabularnewline
$2^{+}$  & 1.0812  & 1.0589  & 1.2740 \tabularnewline
$4^{+}$  & 2.1096  & 1.0588  & 1.8584 \tabularnewline
$6^{+}$  & 2.8883  & 1.0588  & 2.3929 \tabularnewline
$8^{+}$  & 3.2302  & 1.0571  & 2.5125 \tabularnewline
$10^{+}$  & 4.8815  & 1.0464  & 3.2142 \tabularnewline
$12^{+}$  & 5.3394  & 0.9670  & 3.1348 \tabularnewline
$14^{+}$  & 5.4031  & 0.6570  & 2.8247 \tabularnewline
$16^{+}$  & 5.2247  & 0.0000  & 2.1678 \tabularnewline
\end{tabular}\end{ruledtabular} 
\end{table}

We see that despite the 0.9451 overlap between CCGI and E(9), the
even-$I$ spectrum for E(9) in which only the $J=9^{+}$ matrix element
is non-zero is drastically different than CCGI. First of all, the
ground state does not have $I=0$, rather it has $I=J_{\text{max}}=16$
and indeed the two spectra seem to have nothing to do with each other.

Let us briefly digress and look at the spectrum for E(9) for its own
sake. It is quite remarkable. The energies of the $I=0,2,4,6$, and
8 states are very close to each other, differing at most by 0.002~MeV
and the $I=10^{+}$ state is 0.012~MeV lower. All six states are
essentially degenerate. Then there is a drop in energy with $I=16^{+}$
becoming the ground state. Such a strange spectrum and this for an
interaction that gives a 0.9467 overlap with a realistic interaction
for the $I=0^{+}$ state.

In the last column of Table~\ref{tab:yrast}, we improve things by
also lowering the $J=0^{+}$ matrix element to the same value as for
$J=9^{+}$, $-2.0000$~MeV. The spectrum is better, with $I=0^{+}$
now the lowest state, but it is far from satisfactory. Even an overlap
exceeding 0.99 does not guarantee overall good results. Clearly all
two-body matrix elements come into play.

As noted above, the eigenfunction components of the lowest $I=0^{+}$
state for the E(9) interaction are $N\langle(jj)^{J_{p}}(jj)^{J_{n}}|(jj)^{9}(jj)^{9}\rangle^{0}$.
It can be shown that the normalization factor is $N=\sqrt{2}$.

For the $I=1^{+}$ states with the E(9) interaction, the secular matrix
is also separable. This is not true for other values of $J_{\text{odd}}$.
If we were to replace $I=0$ by $I=1$ in the expression in the last
paragraph, all the $U9$-$j$ coefficients would vanish. We must make
a different choice. The eigenfunction components of the lowest $I=1^{+}$
state is then exactly given by a single set of $U9$-$j$ coefficients:
$2\langle(jj)^{J_{p}}(jj)^{J_{n}}|(jj)^{9}(jj)^{8}\rangle^{I=1}$.

For states with $I=2$ or higher, the secular matrix is no longer
separable---rather it is a sum of separable terms. The eigenvalue
equation is 
\begin{eqnarray}
\lefteqn{4\sum_{J_{x}}\langle(jj)^{J_{p}}(jj)^{J_{n}}|(jj)^{9}(jj)^{J_{x}}\rangle^{I}\times}\nonumber \\
 &  & \times\sum_{J_{p'}J_{n'}}\langle(jj)^{J_{p'}}(jj)^{J_{n'}}|(jj)^{9}(jj)^{J_{x}}\rangle^{I}X_{J_{p'}J_{n'}}=\nonumber \\
 & = & \lambda X_{J_{p}J_{n}}\,,\label{eq:i2ev}
\end{eqnarray}
 where $\lambda$ is the eigenvalue and $X_{J_{p}J_{n}}$ stands for
the eigenfunction components. For $I=2^{+}$ there are two terms corresponding
to $J_{x}=7$ and 9; for $I=3^{+}$ the values are $J_{x}=6$ and
8, etc.

Despite the complexity of the above equation, there are some surprising
results. The eigenfunction components of the lowest $2^{+}$ state
are numerically extraordinarily close to the single $U9$-$j$ symbols
$\sqrt{2}\langle(jj)^{J_{p}}(jj)^{J_{n}}|(jj)^{9}(jj)^{9}\rangle^{I=2}$.
Furthermore, the next $2^{+}$ state has also components exceedingly
close to $2\langle(jj)^{J_{p}}(jj)^{J_{n}}|(jj)^{9}(jj)^{7}\rangle^{I=2}$.
This is by no means obvious because, as mentioned above, the interaction
involves a sum of two separable terms corresponding to $J_{x}=7$
and 9.

We can explain this result by performing a calculation of the overlap
of the two $U9$-$j$'s of the last paragraph. We restrict the sum
to even $J_{p}$ and even $J_{n}$. We first note schematically 
\begin{eqnarray}
\lefteqn{4\sum_{\text{even }J_{p},J_{n}}=\sum(1+(-1)^{J_{p}})(1+(-1)^{J_{n}})=}\nonumber \\
 & = & \sum+\sum(-1)^{J_{p}}+\sum(-1)^{J_{n}}+\sum(-1)^{J_{p}+J_{n}}\label{eq:sch}
\end{eqnarray}

The first term vanishes because of Eq.~(\ref{eq:orth}). In the last
term one of the $U9$-$j$'s has two rows that are the same, which
means that the only non-vanishing terms in the sum have ($J_{p}+J_{n}$)
even. Thus, the last term is the same as the first term---zero. The
two middle terms are the same, so we get the overlap of the two $U9$-$j$'s
to be \begin{subequations} 
\begin{eqnarray}
\mathop{\sum_{\text{even }}}_{J_{p},J_{n}} & = & \frac{1}{2}\mathop{\sum_{\text{even }}}_{J_{p},J_{n}}(-1)^{J_{p}}\langle(jj)^{J_{p}}(jj)^{J_{n}}|(jj)^{9}|(jj)^{9}\rangle^{I=2}\times\nonumber \\
 &  & \times\langle(jj)^{J_{p}}(jj)^{J_{n}}|(jj)^{9}(jj)^{7}\rangle^{I=2}=\\
 & = & -\frac{1}{2}\langle(jj)^{9}(jj)^{9}|(jj)^{9}(jj)^{7}\rangle^{I=2}\,.\label{eq:i2}
\end{eqnarray}
 \end{subequations} We obtain the above by using the orthogonality
relations for $9j$-symbols as shown in Eqs.~(\ref{eq:orth}) and~(\ref{eq:u9j-rel2}).

Using similar arguments, one can show that the normalization for the
$|[pn(9)pn(9)]^{I=2}\rangle$ state is such that its normalization
factor is 
\begin{eqnarray}
N(9)^{-2} & = & \frac{1}{2}-\frac{1}{2}\langle(jj)^{9}(jj)^{9}|(jj)^{9}(jj)^{9}\rangle^{I=2}=\label{eq:n2-99}\\
 & = & \frac{1}{2}-\frac{1}{2}0.00001209813=0.499993950935\nonumber 
\end{eqnarray}

For the $|[pn(9pn(7)]^{I=2}\rangle$ state, we obtain 
\begin{eqnarray}
N(7)^{-2} & = & \frac{1}{4}-\frac{1}{2}\langle(jj)^{9}(jj)^{7}|(jj)^{9}(jj)^{7}\rangle^{I=2}=\label{eq:n2-97}\\
 & = & \frac{1}{4}+\frac{1}{2}0.00075253477=0.250376267385\nonumber 
\end{eqnarray}

To get this latter result, we use the following relationship 
\begin{equation}
\sum_{J_{p},J_{n}}(-1)^{(J_{p}+J_{n})}\left|\langle(jj)^{9}(jj)^{7}|(jj)^{J_{p}}(jj)^{J_{n}}\rangle^{I=2}\right|^{2}=0
\end{equation}

From Eqs.~(\ref{eq:n2-99}) and (\ref{eq:n2-97}), we find that the
normalizations are 1.414222 and 1.998497, only slightly different
than $\sqrt{2}$ and 2 respectively. Therefore, we obtain that the
term in Eq.~(\ref{eq:i2}) is exceedingly small for the $g_{9/2}$
shell, namely 0.00009113 and, if we include the exact normalization
factors, we get 0.00025756.

In lower shells the deviations are larger. To see the trend, we give
in Table~\ref{tab:u9j-sh} the value of the $U9$-$j$ symbol of
Eq.~(\ref{eq:i2}) with 9 replaced by $J_{\text{max}}$ and 7 by
$J_{\text{max}}-2$ for various shells. The last column uses approximate
normalization factors 2 and $\sqrt{2}$. Clearly the overlap approaches
zero in the large-$j$ limit. As we go from one shell to the next,
the value of the overlap drops by at least a factor of 10. A study
of the large-$j$ behaviour of 9-$j$ symbols has been performed by
L.~Yu and R.G.~Littlejohn~\cite{yl11}.

\begin{table}[htb]
 \caption{\label{tab:u9j-sh} Value of coupling $U9$-$j$ symbols for various
shells.}

\begin{ruledtabular} %
\begin{tabular}{ccc}
$j$  & $U9$-$j$  & overlap of Eq.~(\ref{eq:i2}) \tabularnewline
\hline 
$p_{3/2}$  & $-0.1800$  & 0.2546 \tabularnewline
$d_{5/2}$  & $-0.021328$  & 0.03016 \tabularnewline
$f_{7/2}$  & $-0.002074$  & 0.002933 \tabularnewline
$g_{9/2}$  & $-0.0001822$  & 0.0002577 \tabularnewline
$h_{11/2}$  & $-0.00001502$  & 0.00002174 \tabularnewline
$i_{13/2}$  & $-0.000001185$  & 0.000001676 \tabularnewline
\end{tabular}\end{ruledtabular} 
\end{table}

We can see in Table~\ref{tab:j2} that the results for matrix diagonalization
for both $I=2^{+}$ states yield wave function components which are
very close to the normalized $U9$-$j$ coefficients. In fact, they
are so close that one could wonder if they are exactly the same. But
they are not. As seen in Eq.~(\ref{eq:i2}), the two $U9$-$j$ sets
corresponding to $J_{x}=9$ and $J_{x}=7$ are very nearly orthogonal,
but not quite.

\begin{table}[htb]
 \caption{\label{tab:j2} Comparison for the first two $I=2^{+}$ states of
the matrix diagonalization with the E(9) interaction and with normalized
$U9$-$j$ components. We give the energy in MeV in the second row.}

\begin{ruledtabular} %
\begin{tabular}{ccccc}
$[J_{p}\,,J_{n}]$  & E(9)  & $U9$-$j$  & E(9)  & $U9$-$j$ \tabularnewline
 & 1.069  &  & 3.0558  & \tabularnewline
\hline 
$[0\,,2]$  & 0.5334  & 0.5338  & 0.1349  & 0.1351 \tabularnewline
$[2\,,2]$  & $-0.4707$  & $-0.4708$  & 0.5569  & 0.5567 \tabularnewline
$[2\,,4]$  & 0.3035  & 0.3035  & 0.3188  & 0.3189 \tabularnewline
$[4\,,4]$  & $-0.1388$  & $-0.1390$  & 0.6300  & 0.6299 \tabularnewline
$[4\,,6]$  & 0.0531  & 0.0531  & 0.1320  & 0.1320 \tabularnewline
$[6\,,6]$  & $-0.0137$  & $-0.0138$  & 0.1350  & 0.1350 \tabularnewline
$[6\,,8]$  & 0.0025  & 0.0025  & 0.0114  & 0.0114 \tabularnewline
$[8\,,8]$  & $-0.0003$  & $-0.0003$  & 0.0052  & 0.0052 \tabularnewline
\end{tabular}\end{ruledtabular} 
\end{table}

It turns out that all the other lowest even-$I$ states have eigenfunctions
close although not exactly equal to $\sqrt{2}\langle(jj)^{J_{p}}(jj)^{J_{n}}|(jj)^{9}(jj)^{9}\rangle^{I}$.
In Table~\ref{tab:9j} we compare, as an example, the wave function
of the $I=8^{+}$ state. In the second column, we give the single
$U9$-$j$ symbols (normalized) and in the third column we give results
of diagonalizing the E(9) interaction. Since the coefficient $[J_{p},J_{n}]$
is the same as $[J_{n},J_{p}]$, we list only one of them. The overlap
of the two wave funtions is 0.9998.

\begin{table}[htb]
 \caption{\label{tab:9j} Comparing the wave functions of a single $U9$-$j$
symbol with $J_{x}=9$ with a full diagonalization of E(9) for the
lowest $I=8^{+}$ state in $^{96}$Cd.}

\begin{ruledtabular} %
\begin{tabular}{ccc}
$[J_{p}\,,J_{n}]$  & $U9$-$j$  & E(9) \tabularnewline
\hline 
$[0\,,8]$  & 0.0630  & 0.0644 \tabularnewline
$[2\,,6]$  & 0.4299  & 0.4271 \tabularnewline
$[2\,,8]$  & $-0.0522$  & $-0.0513$ \tabularnewline
$[4\,,4]$  & 0.7444  & 0.7456 \tabularnewline
$[4\,,6]$  & $-0.1803$  & $-0.1729$ \tabularnewline
$[4\,,8]$  & 0.0256  & 0.0280 \tabularnewline
$[6\,,6]$  & 0.0521  & 0.0657 \tabularnewline
$[6\,,8]$  & $-0.0076$  & $-0.0012$ \tabularnewline
$[8\,,8]$  & 0.0011  & 0.0047 \tabularnewline
\end{tabular}\end{ruledtabular} 
\end{table}

\subsection{Isospin considerations}

The E(9) interaction is charge independent and therefore one can assign
a definite isospin to a non-degenerate state. A wave function for
two protons and two neutrons in a single $j$-shell can be written
as 
\begin{equation}
\Psi^{I}=\sum_{J_{p},J_{n}}X_{J_{p}J_{n}}|[pp(J_{p})nn(J_{n})]^{I}\rangle\,,
\end{equation}
 where $X_{J_{p}J_{n}}$ is the probability amplitude that the protons
are in the state $J_{p}$ and the neutrons in the state $J_{n}$.
Again, $I$ is the total angular momentum.

If one uses a charge-independent interaction, then for cases where
the number of valence protons equals the number of valence neutrons,
the states fall into two classes. In the first, $X_{J_{p}J_{n}}=X_{J_{n}J_{p}}$,
while in the second class $X_{J_{p}J_{n}}=-X_{J_{n}J_{p}}$. In our
case of two protons and two neutrons, for even $I$ the states in
the first class must have even isospin, $T=0$ or 2; in the second
class, the states have $T=1$. For odd $I$ the states in the first
class must have odd isospin, $T=1$; in the second class, $T=0$ or
2.

For $I=0$ and $I=1$, each wave function component has $J_{p}=J_{n}$,
so all these states belong to class 1. For $I=0$, the states must
have isospins $T=0$ or $T=2$; for $I=1$, all states must have isospin
$T=1$.

We can now make an association of isospin with the quantum number
$J_{x}$ for those states for which this is an exact or reasonably
approximate good quantum number. If we interchange two rows (or two
columns) of a 9-$j$ symbol, the result is the same 9-$j$ symbol
multiplied by a phase factor $(-1)^{s}$, where $s$ is the sum of
all nine angular momenta in the 9-$j$ symbol. If two rows are identical
and $s$ is odd, the 9-$j$ will be equal to zero. This leads to the
result that for even total angular momentum $I$, we must have $J_{x}$
odd for states with even isospin, i.e. $T=0$ or 2, whilst $J_{x}$
must be even for $T=1$ states. For odd $I$, $J_{x}$ must be odd
for $T=0$ and $T=2$ states and even for $T=1$ states.

In the single $j$-shell there are no $I=0$, $T=1$ states, so $J_{x}$
must be odd. For $I=1$, all states have $T=1$, so $J_{x}$ is even.
For states with $I=2$ and 3, and indeed for most other states, one
can have all possible isospins ($T=0,1,2$), so one must make separate
analyses with even and odd values of $J_{x}$.

It is not a priori clear if the above mentioned $I=0$ state with
$J_{x}=9$ has isospin $T=0$ or 2; likewise the above two $I=2^{+}$
states. One can show that all these states have isospin $T=0$. The
reason is that a $T=2$ state will have an absolute energy $E=0$,
since it is a double analog of a state with four identical nucleons.
For the latter system, all pairs have $T=1$, whereas the E(9) interaction
acts only for two nucleons in a $T=0$ state. Four identical nucleons
do not see any interaction. The $I=0$ and both $I=2$ states above
are non-degenerate and have finite absolute energies. For example,
as will be shown in a later section on degeneracies, the absolute
energy of the lowest $I=0$ state is $2V(9)$. Indeed this $I=0$
state and the two $I=2^{+}$ states above all have isospin $T=0$.

The E(9) interaction also yields a $T=1$, $I=2$ non-degenerate state
with components $2\langle(jj)^{9}(jj)^{8}|(jj)^{J_{p}}(jj)^{J_{n}}\rangle^{2}$.
This is a pure state---it does not mix with any other $T=1$ state.
This is because there is only one way of forming an $I=2$, $T=1$
state from $U9$-$j$ symbols, i.e only one possible $J_{x}=8$. With
$J_{x}=6$ combined with $J=9$, we cannot get $I=2$.

For $I=3$, $T=0$ there is also a pure state $2\langle(jj)^{9}(jj)^{7}|(jj)^{J_{p}}(jj)^{J_{n}}\rangle^{3}$.
This wave function changes sign under the interchange of $J_{p}$
and $J_{n}$. It cannot admix with a state with $J_{x}=8$ or $J_{x}=6$
because those states have $T=1$.

\subsection{Comparison with realistic interactions}

At the beginning of Section~\ref{sec:results} we compared the overlaps
between the $I=0$ eigenstates of various interactions with a realistic
one (CCGI~\cite{ccgi12}). Now we want to extend the comparison to
all the yrast states of the two most relevant interactions in this
study: E(9) and E(0,9). Thus, in Table~\ref{tab:comp} we present
the two sets of overlaps with the realistic interaction for the lowest
even-$I$ states from $I=0$ to $I=16$ for the $2p$-$2n$ system,
e.g. $^{96}$Cd. In the second column we have the overlap $\langle\psi_{\text{E(9)}}|\psi_{U9j}\rangle$
between the eigenstates of the E(9) interaction and the normalized
$U9$-$j$ coefficients $\sqrt{2}\langle(jj)^{9}(jj)^{9}|(jj)^{J_{p}}(jj)^{J_{n}}\rangle^{I}$.
In the third column we have the overlap $\langle\psi_{\text{E(9)}}|\psi_{\text{CCGI}}\rangle$
between eigenstates of the E(9) interaction and eigenstates of the
CCGI interaction. In the fourth column we use the interaction E(0,9).
Adding the attractive $J=0$ pairing interaction only affects states
from $I=0$ to $I=8$.

\begin{table}[htb]
 \caption{\label{tab:comp} Overlaps between the eigenstates of the E(9) and
E(0,9) interactions with the more realistic CCGI interaction~\cite{ccgi12}.
We also give the overlap between the eigenstates of E(9) and normalized
$U9$-$j$ coefficients.}

\begin{ruledtabular} %
\begin{tabular}{cccc}
$I$  & $\langle\psi_{\text{E(9)}}|\psi_{U9j}\rangle$  & $\langle\psi_{\text{E(9)}}|\psi_{\text{CCGI}}\rangle$  & $\langle\psi_{\text{E(0,9)}}|\psi_{\text{CCGI}}\rangle$ \tabularnewline
\hline 
0  & 1  & 0.9451  & 0.9944 \tabularnewline
2  & 0.99996  & 0.9829  & 0.9904 \tabularnewline
4  & 0.99995  & 0.9144  & 0.9773 \tabularnewline
6  & 0.99950  & 0.6795  & 0.9361 \tabularnewline
8  & 0.99977  & 0.2375  & 0.9858 \tabularnewline
10  & 0.96377  & 0.6830  & 0.6830 \tabularnewline
12  & 0.98125  & 0.9944  & 0.9944 \tabularnewline
14  & 0.95007  & 0.9967  & 0.9967 \tabularnewline
16  & 1  & 1  & 1 \tabularnewline
\end{tabular}\end{ruledtabular} 
\end{table}

We see that the overlaps between E(9) and $U9$-$j$ for $J=0$ and
16 are 1. The latter is trivial because there is only one configuration
for $I=16$: $J_{p}=8$, $J_{n}=8$. The $I=0$ result is explained
by the fact that the Hamiltonian is separable in this case. The overlaps
for $I=2$ and 4 are very close to 1. In general, the overlaps are
very high, so that the ansatz $|[pn(9)pn(9)]^{I}\rangle$ is very
good but not 100\%. It appears that in the large-$j$ limit it would
be exact.

In the next column we give the overlaps of the simple E(9) interaction
with the more realistic CCGI interaction. We get overlaps bigger than
$0.9$ for $I=0,2,4,12,14$, and 16. Again $I=16$ is a trivial case.
But why does the overlap that goes down suddenly rise up for $I=12$
and 14? It must be because the interaction E(0), i.e. $J=0$ pairing,
does not come into play here. We see that for $I=6$, 8, and 10 the
simple ansatz does not yield a satisfactory wave function.

In the last column we see that the poor overlaps for $I=6$ and 8
can be improved significantly by adding a $J=0$ pairing term to the
interaction. We now have $V(0)=V(9)$ (attractive). The overlap for
$I=6$ increases from 0.6795 to 0.9361 and for $I=8$ from 0.2375
to 0.9858. The overlaps for $I=0,2$, and 4 are also improved. For
$I=10$ and higher, one cannot have any pair coupled to zero, so the
E(0,9) interaction gives the same result as E(9).

\subsection{Degeneracies}

With the E(9) interaction, we get several degenerate states with an
absolute energy zero. In some detail, for $I=0^{+}$ there are five
states, three with isospin $T=0$ and two with $T=2$. There is one
non-degenerate state at an energy $2V(9)$ ($V(9)$ is negative).
The other four $I=0^{+}$ states have zero energy. For $I=1^{+}$
all states have isospin $T=1$. There is a single non-degenerate state
at $V(9)$, the other three have zero energy. For $I=2^{+}$ there
are twelve states---six have $T=0$, four have $T=1$, and two have
$T=2$. There are two non-degenerate $T=0$ states with approximate
energies $2V(9)$ and $V(9)$ respectively, and one non-degenerate
$T=1$ state with energy $V(9)$. The other nine states have zero
energy. To understand this, take a wave function 
\[
|\Psi^{\alpha}\rangle=\sum_{J_{p},J_{n}}{C^{\alpha}(J_{p},J_{n})|[pp(J_{p})nn(J_{n})]^{I}\rangle}
\]
 and the corresponding energies $E^{\alpha}=\langle\Psi^{\alpha}|H|\Psi^{\alpha}\rangle$.
Consider the sum $\sum_{\alpha}{E^{\alpha}}$. We have 
\begin{equation}
\sum_{\alpha}{C^{\alpha}(J_{p},J_{n})C^{\alpha}(J_{p'},J_{n'})}=\delta_{J_{p},J_{p'}}\delta_{J_{n},J_{n'}}
\end{equation}
 Thus 
\begin{eqnarray}
\lefteqn{\sum_{\alpha}{E^{\alpha}}=\sum_{J_{p}J_{n}}{\langle[pp(J_{p})nn(J_{n})]^{I}|H|[pp(J_{p})nn(J_{n})]^{I}\rangle}=}\\
 & = & 4V(9)\mathop{\sum_{J_{p}J_{n}}}_{\text{even}}{\sum_{J_{x}}{\left|\langle(jj)^{J_{p}}(jj)^{J_{n}}|(jj)^{9}(jj)^{J_{x}}\rangle^{I}\right|^{2}}}\label{eq:sume}
\end{eqnarray}

This expression does not depend on the detailed wave functions. Referring
to Eqs.~(\ref{eq:n2-99}) and (\ref{eq:n2-97}) and neglecting the
very small correction terms, we see that $N^{-2}$ is equal to $1/2$
for $J_{x}=9$ and to $1/4$ for all other $J_{x}$. Basically then
Eq.~(\ref{eq:sume}) becomes $4V(9)\sum N(J_{x})^{-2}$. Hence we
obtain $\sum_{\alpha}{E^{\alpha}}=2V(9)$ for $I=0$, $V(9)$ for
$I=1$, and $4V(9)$ for $I=2$. But we can alternately show, using
the explicit wave functions, that for $I=0$ the energy of the lowest
state is $2V(9)$. Hence, all the other states must have zero energy.
A similar story for $I=1$. The $I=2$ state is a bit more complicated
because of the coupling between two states, however small it is. Still
one can work it through and see that the $4V(9)$ energy is exhausted
by the two $T=0$ and the one $T=1$ non-degenerate states.

For $I=0$ we have two $T=0$ and two $T=2$ states, all degenerate.
One can remove the degeneracies of $T=0$ and $T=2$ by adding to
the Hamiltonian an interaction $b\; t(i)\cdot t(j)$. This will not
affect the wave functions of the non-degenerate states but will shift
the $T=2$ states away from the formerly degenerate $T=0$ states.

\section{Closing remarks}

In closing, we note that the subject of $J_{\text{max}}$ pairing
is currently a very active field. Besides the work of Qi et al.~\cite{qetal11},
there are related works by Zerguine and Van Isacker~\cite{zi11},
Cederwall et al.~\cite{cetal11} and Xu et al.~\cite{xetal12}.
The topic of $J$-pairing interactions has also been addressed by
Zhao and Arima~\cite{za05}. In this work we expand on our 2003 work~\cite{mrzs03}
by renoting that the Hamiltonian matrix for a $2p$-$2n$ system for
$I=0^{+}$ states in a single $j$-shell is separable for a simple
interaction which is non-zero only for a single odd angular momentum.
This leads to an eigenfunction with components proportional to a single
set of unitary 9-$j$ symbols. We apply this to the $J=J_{\text{max}}$
interaction. The single set of $U9$-$j$ components form the eigenfunction
not only for the lowest $I=0^{+}$ state, but also of the lowest $I=1^{+}$
state and, to a surprisingly excellent approximation, for the lowest
two $I=2^{+}$ states. A single set of $U9$-$j$ coefficients yields
a good approximation for all yrast even-$I$ states. We also note
that the non-degenerate states have good isospin and we show how to
assign this quantum number. We have found a quantum number $J_{x}$
(see Eq.~(\ref{eq:i2ev})) which can, either exactly or approximately,
help classify some of the non-degenerate states. We have shown how
to determine the number of degenerate states. We have compared our
results of the E(9) interaction with those of E(0,9) in which $J=0$
and $J=2j$ matrix elements are equally attractive. Also comparison
with a realistic interaction have been made. With E(9) alone, only
$I=12$, 14, and (trivially) 16 have strong overlaps with the realistic
interaction, but the situation is dramatically improved with E(0,9).
Lasty, we feel that just as there has been an intensive study of the
$J=0$ {}``pairing interaction'' with many positive benefits, so
should we as a counterpoint make an intensive study of the E($J_{\text{max}}$)
interaction as well as other schematic interactions. We already have
the fascinating result of the closeness, shown in Table~\ref{tab:comp},
between $\psi_{\text{E(9)}}$ and $\psi_{U9j}$. Perhaps other interesting
results will be found later. At the same time, we must also keep a
reality check on both the $J=0$ and $J=J_{\text{max}}$ {}``pairing
interactions'' and this we have done by displaying the results in
the other columns of the same table.
\begin{acknowledgments}
We are indebted to Ben Bayman for many useful comments and insights. \end{acknowledgments}

\end{document}